\documentclass[usenatbib]{mn2e}

\usepackage{graphicx}
\usepackage{amssymb}
\usepackage{epsfig}

\title[SXP327]{An optical \& X-ray study of the counterpart to the SMC X-ray binary pulsar system SXP327}

\author[M.J. Coe et al.]{M. J.~Coe$^{1}$,
 M. ~Schurch$^{1}$, R.H.D. ~Corbet$^{2}$,
J. Galache$^{3}$, \and V.A. ~McBride$^{1}$, L.J. Townsend $^{1}$, A. Udalski$^{4}$\\
$^{1}$ School of Physics and Astronomy, University of Southampton, SO17
1BJ, UK\\
$^{2}$ University of Maryland, Baltimore County, Mail Code 662, NASA Goddard Space Flight Center, Greenbelt, MD 20771, USA \\
$^{3}$ Harvard-Smithsonian Center for Astrophysics, 60 Garden Street,
Cambridge, MA 02138, USA. \\
$^{4}$ Warsaw University Observatory, Aleje Ujazdowskie 4, 00-478 Warsaw, Poland \\
}

\begin{document}

\date{26 Mar 2008}

\pagerange{\pageref{firstpage}--\pageref{lastpage}} \pubyear{2002}

\maketitle

\label{firstpage}

\begin{abstract}

Optical and X-ray observations are presented here of a newly reported X-ray transient system in the Small Magellanic Cloud. The data reveal many previously unknown X-ray detections of this system and clear evidence for a 49.995d binary period. In addition, the optical photometry show recurring outburst features at the binary period which may well be indicative of the neutron star interacting with a circumstellar disk around a Be star.

\end{abstract}


\section{Introduction and background}

The Be/X-ray systems represent the largest sub-class of massive X-ray
binaries.  A survey of the literature reveals that of the 115
identified massive X-ray binary pulsar systems (identified here means
exhibiting a coherent X-ray pulse period), most of the systems fall
within this Be counterpart class of binary.  The orbit of the Be star
and the compact object, presumably a neutron star, is generally wide
and eccentric.  X-ray outbursts are normally associated with the
passage of the neutron star close to the circumstellar disk (Okazaki
\& Negueruela 2001). A detailed review of the X-ray properties of such systems may be found in Sasaki et al. (2003) and a review of the more general properties can be found in Coe et al. (2000).

The source that is the subject of this paper was first identified by Laycock et al (2008) as part of their deep Chandra study of one region in the Small Magellanic Cloud (SMC). It has a pulse period of 327s so it is designated here as SXP327 following the naming convention of Coe et al. (2005). In this paper over 10 years of optical photometry data of the optical counterpart are presented which show clear evidence for a binary period. These data are combined with $\sim$10 years of Rossi X-ray Timing Explorer (RXTE) data which show several earlier outbursts from this system and also confirm the binary period seen in the optical.

\section{Optical data}

\begin{figure}
\begin{center}
\includegraphics[width=60mm,angle=-90]{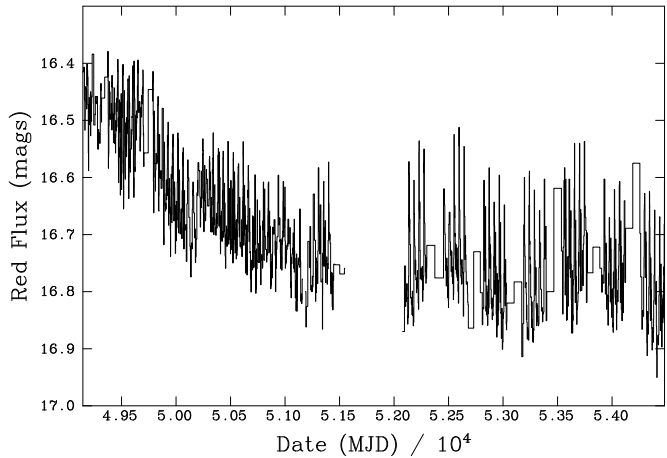}
\caption{The combined optical lightcurves from MACHO (left) and OGLE (right). The
MACHO red data have been arbitrarily rescaled to match with
the OGLE I-band data.}
\label{fig:merged}
\end{center}
\end{figure}

The X-ray position for SXP327 is reported by Laycock et al (2008) as
  $00^{\rm h}52^{\rm m} 52.3^{\rm s},-72^\circ 17'15.4''$
and a 95\% error circle of radius $1.1''$. A visual check of the digitized sky image for this position reveals a clear optical counterpart consistent with this position. This object can be found in the OGLE III data base as SMC101.4 25097, and also in the MACHO data base as object 207.16147.60. The OGLE III data indicate an I magnitude of $\sim$16.7. The two combined data sets are shown in Figure~\ref{fig:merged}. Since the MACHO data are not readily converted to Johnson magnitudes the red data shown in Figure~\ref{fig:merged} have been arbitrarily adjusted by the addition of an offset of 24.5 to bring it into approximate alignment with the OGLE III I-band results where they join.

\begin{figure}
\begin{center}
\includegraphics[width=60mm,angle=-90]{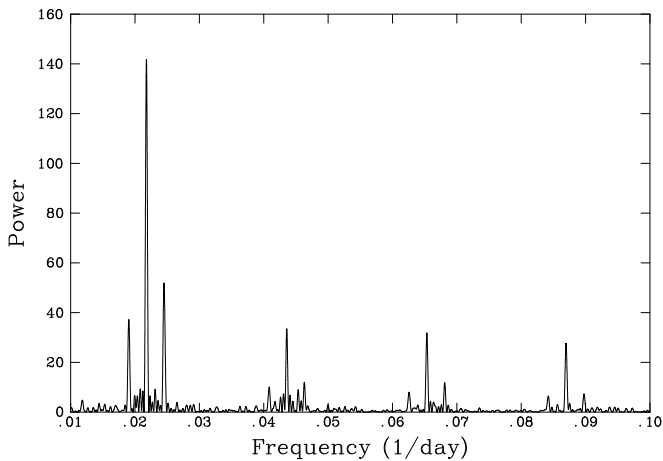}
\caption{The Lomb-Scargle power spectrum obtained from the OGLE III data. The main peak is at a period of 45.995d. The side peaks are the beats between the main period and the annual sampling interval.}
\label{fig:power}
\end{center}
\end{figure}

It is immediately apparent from the raw data that there exists a strong optical modulation in the photometry. So the combined data set were initially detrended with a 3rd order polynomial and then analyzed with a Lomb-Scargle routine to search for periodicities. A very strong clear peak was seen in the power spectrum at 45.995d - see Figure~\ref{fig:power} - with the other strong peaks being exact harmonics of this period. The side peaks are due to the beating of the true period with the annual sampling interval.

\begin{figure}
\begin{center}
\includegraphics[width=80mm,angle=-0]{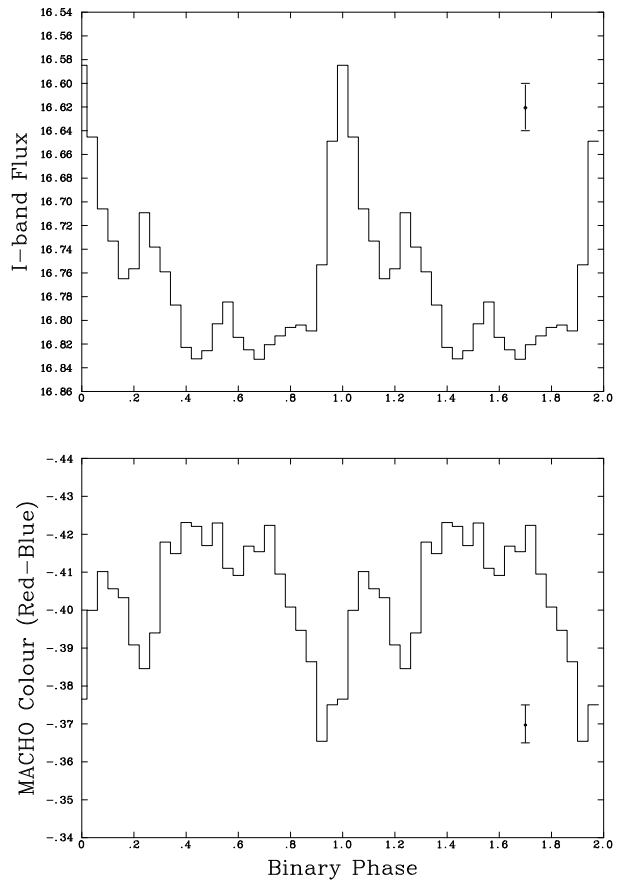}
\caption{Top panel : the OGLE III data set folded at the proposed binary
period of 45.99d. Lower panel : the MACHO colour data (red - blue)
folded at the same period. In each panel the typical size of the uncertainty on the points is indicated.}
\label{fig:foldcol}
\end{center}
\end{figure}

Consequently the data were folded at this period to determine the average profile - see Figure~\ref{fig:foldcol} - revealing a strong outburst peak of full-width half-maximum of $\sim$0.2 in phase, or $\sim$9d. Two other probable peaks are also revealed at phases 0.25 and 0.55 after the main peak. Since the MACHO data provide both red and blue data coverage, the difference of each pair of measurements was calculated and then the resulting colour data were folded at the same period. The result is shown in the lower panel of Figure~\ref{fig:foldcol}. Again it is possible to clearly see the presence of two significant features in the folded data correlating accurately with the two largest outburst peaks. There is also some indication of a matching colour change close to the phase of the third peak.

From the OGLE III data set the ephemeris for outbursts is determined to be:

$T_{outburst}$ = (53149.0$\pm$1.0) + n(45.99$\pm$0.03) MJD

\section{X-ray data}

\begin{figure*}
\begin{center}
\includegraphics[width=120mm,angle=90]{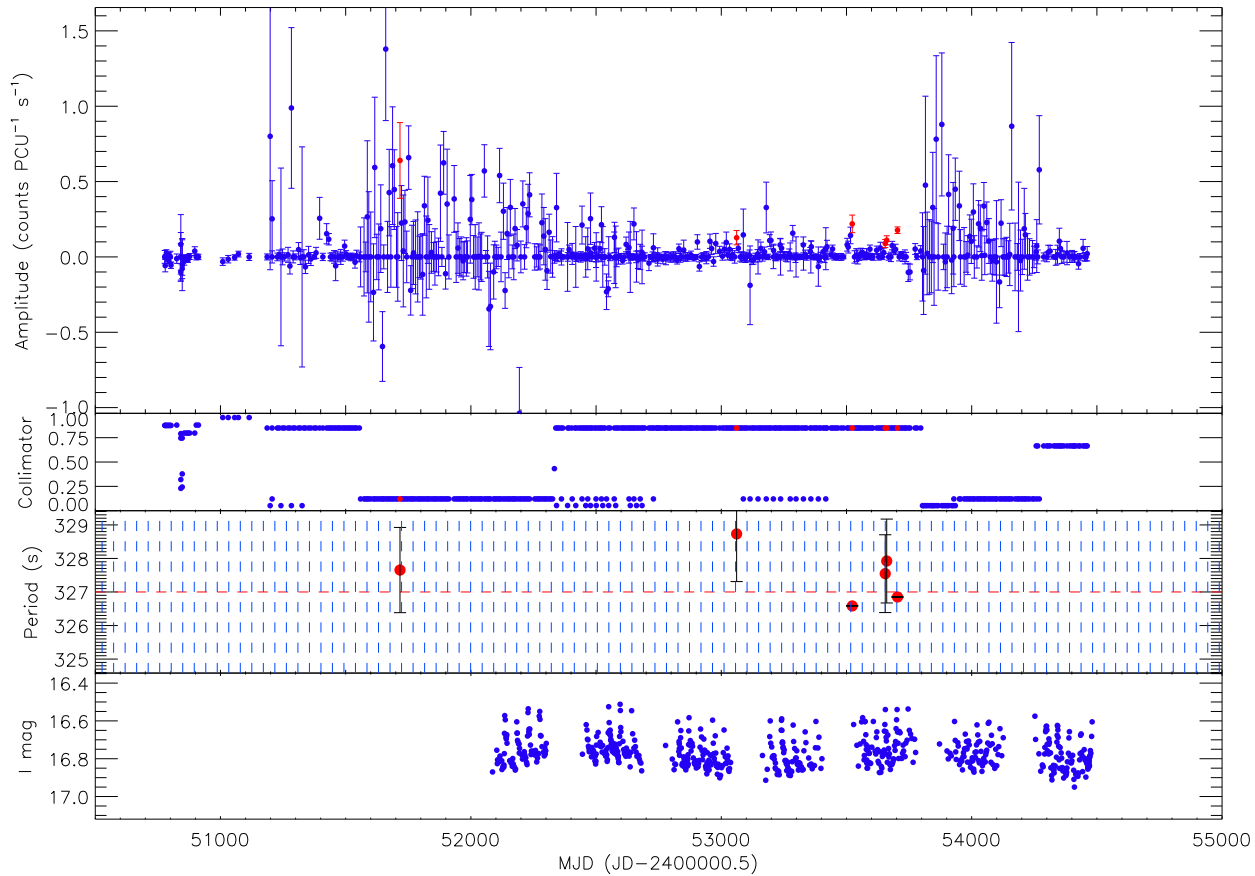}
\caption{More than 10 years of RXTE monitoring data of the SMC searched for evidence of 327s pulsations. Top panel : RXTE Lomb-Scargle pulse amplitude; second panel : the RXTE collimator response to SXP327; third panel : pulse period detected; fourth and lowest panel : the I-band magnitude determined from the OGLE III data. The vertical lines show the phase 0.0 epochs according to the ephemeris quoted at the end of Section 2.}
\label{fig:sxom}
\end{center}
\end{figure*}

Since the SMC has been subject to extensive monitoring using RXTE (Laycock et al, 2005, Galache et al, 2008) it was possible to use these data to search for evidence of previous X-ray detections of SXP327. As a result, six other detections of this pulsar were found - see Table~\ref{tab:obs}. The totality of the RXTE coverage is shown in Figure~\ref{fig:sxom}; the XTE pulsed flux is measured in the 3-10keV band. There was no confusion with another object of similar period, SXP323, since the RXTE data are more than adequate to separate the two periods.

X-ray luminosities were calculated based on the strength of the pulse amplitude and a powerlaw spectrum with a spectral index of 1. Assuming a distance of 60kpc for the SMC, that the flux is 100\% pulsed, and using PIMMS v3.9e (http://heasarc.nasa.gov/Tools/w3pimms.html), the resulting values lie in the range $L_{X}$=(0.8-2.0)$\times10^{36}$ erg/s, not uncharacteristic of Type I outbursts from these systems.
The pulse fraction measured from the folded lightcurve as ($F_{max}-F_{min})/(F_{max}+F_{min})$ comes out as 8.5\% which is very low. One reason for this low value could be the contribution of other non-pulsing SMC sources to the base-line signal. However, if we assume a pulsed fraction of $\sim$10\% then the X-ray luminosity rises to $\sim$1$\times10^{37}$ erg/s.

The most significant Proportional Counter Array (PCA) detections of SXP327 all occurred when other pulsars were simultaneously detected in the field of view. For this reason it was not possible to extract a time-averaged PCA spectrum of SXP327 uncontaminated by emission from other sources and measure the X-ray luminosity from SXP327 in that manner.

\begin{table}
\begin{center}
\caption{RXTE X-ray detections of SXP327}
\label{tab:obs}
\begin{tabular}{cccc} \hline
MJD Date & Calendar date & Period detected(s) & Binary phase \\
&&&\\
51716.97 &21 Jun 2000 &327.7$\pm$1.27  &0.13 \\
53061.28 &26 Feb 2004 &328.7$\pm$1.42  &0.91 \\
53522.95 &01 Jun 2005 &326.6$\pm$0.10  &0.13 \\
53654.69 &11 Oct 2005 &327.5$\pm$1.16  &0.00 \\
53660.67 &17 Oct 2005 &327.9$\pm$1.25  &0.12 \\
53703.77 &29 Nov 2005 &326.9$\pm$0.22  &0.07 \\

\hline
\end{tabular}
\end{center}
\end{table}

The phase of the X-ray detections were then determined using the ephemeris quoted above and the results are also included in Table~\ref{tab:obs}. From these numbers it is clear that the X-ray outbursts are strongly correlated with the optical outbursts. It is worth noting that most of the RXTE observations have been carried out at 1 week intervals and this would correspond to a phase interval of 0.15. Therefore the RXTE detections do not necessarily mark the peak of each outburst.
Nonetheless, the results unequivocally confirm that this is the correct optical counterpart to SXP327. It also confirms that the optical period is almost certainly the binary period of the system with the X-ray outbursts occurring around the time of periastron passage of the neutron star - i.e. confirming that they are Type I outbursts.

From the position quoted by Laycock et al. (2008)it is possible that this is the same source as the one identified as no:45 in the catalogue of Sasaski, Haberl \& Pietsch (2000) from ROSAT HRI data. Though the positions differ by $\sim$13" (which is much larger than their quoted positional error of 3.9") there is no obvious optical object at their position. This ROSAT source has no special characteristics listed in their paper.

\section{Discussion}

\begin{figure}
\begin{center}
\includegraphics[width=50mm,angle=-90]{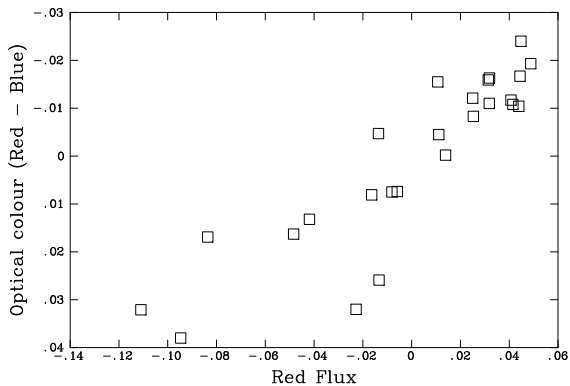}
\caption{Optical colour (MACHO red mag - blue mag) versus optical flux (merged and detrended MACHO red and OGLE III I-band magnitudes).}
\label{fig:colflux}
\end{center}
\end{figure}

SXP327 is an exceptional member of the SMC X-ray binary pulsar systems in that it shows a very strong optical modulation at the binary period. Another source, SXP46.6, has also recently been shown by McGowan et al. (2008) to exhibit optical flaring at the same phase as X-ray outbursts, but not in the same strong and consistent manner as SXP327. Those authors discuss the probable cause of this phenomenon as lying in the periodic disturbance of the Be stars circumstellar disk. At the time of periastron passage Okazaki \& Negueruela (2001) have shown that the disk can be perturbed from its stable, resonant state with a resulting increase in surface area and, hence, optical brightness. What is very unusual about this system, SXP327, is that there is not one, but at least two outbursts every binary cycle at phases 0.0 and 0.25 (i.e. separated by about 11d). In addition, the average profile seems to also show a third peak at phase 0.55 - close to apastron. Figure~\ref{fig:foldcol} shows that the colours of the system reflect the optical brightness and this is made even clearer if the folded flux values are plotted against the folded colour values - see Figure~\ref{fig:colflux}. It is obvious from this figure that the correlation between colour and flux occurs throughout the binary cycle even though it is most prominent at the time of the outbursts. The direction of the correlation is to make the system more bluer when brighter - perhaps an indication of X-ray heating contributing to the colour changes.

\begin{figure}
\begin{center}
\includegraphics[width=70mm,angle=-0]{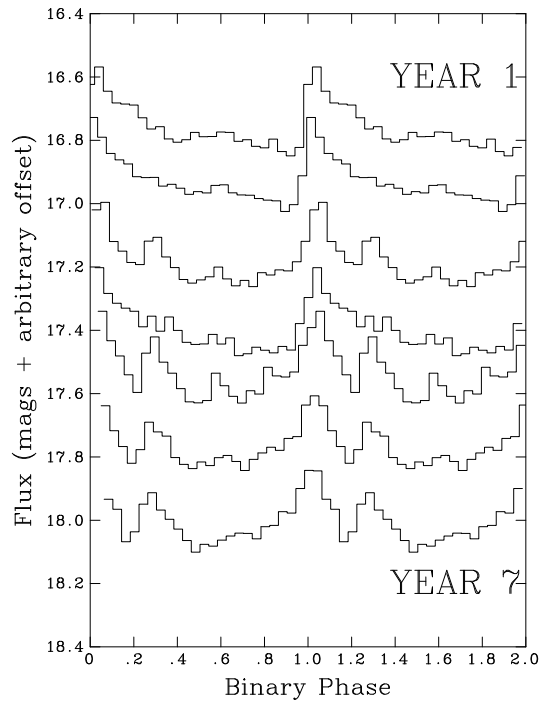}
\caption{Annual variation in the binary profile of the optical photometry obtained from folding the OGLE III data.}
\label{fig:stack}
\end{center}
\end{figure}

To explore the optical modulation further, the 7 years worth of OGLE III data were divided into annual samples and folded at the ephemeris quoted above. The results are shown in Figure~\ref{fig:stack}. From this figure it is possible to see that the main outburst has been strongly present in all the data, but the secondary peak has only been present for the last 3-5 years. Furthermore, the outburst profile may be related to the X-ray activity, or vice versa. From Figure~\ref{fig:sxom} it can be seen that during Years 2-5 SXP327 enjoyed equal exposure in the RXTE observations. Yet only Year 5 shows any significant amount of X-ray activity and it is this year where the largest change in the optical profile occurs with the emergence of the clear, strong double peak structure. It is possible that these changes in the optical outburst profile are related to changes in the circumstellar disk size or structure. Unfortunately there are no optical spectral data available to investigate this further, but clearly this would be of great value in the future.

Another SMC pulsar system, SXP46.6, also shows evidence for a secondary peak in its optical outburst profile (see Figure 4 in McGowan et al., 2008) around phase 0.15. So, in contrast to the outburst profiles for most other sources, the burst peaks for SXP327 and SXP46.6 seem to be split into two. Interestingly, a double peaked orbital outburst profile has been seen in the X-ray emission from at least three other pulsar systems: the supergiant binary GX 301-2 (Pravdo et al. 1995), 1E1145.1-6141 (Corbet et al., 2007) and GRO J1944+26 (Wilson et al. 2003). Though these are very different wavebands, it is still possible to draw comparisons between those sources and SXP327 \& SXP46.6. In the previously reported cases, the two outbursts were explained in terms of a misalignment between the circumstellar disc and the orbital plane of the neutron star. This would lead to the neutron star passing close to, or through the disc twice each
orbit, producing a burst each time. As SXP46.6 does not show two X-ray
bursts (there is inadequate X-ray coverage of SXP327 to provide a detailed orbital profile), it is possible that enough material is transferred to the neutron star to allow the X-ray emission to continue at a stable level, perhaps via an accretion disc, during the two passages.

It is worth noting that, in the absence of a full orbital solution, the assumption has been made here that the first and largest optical peak corresponds to periastron. Of course, this may not be the case, and the two adjacent optical peaks (phase 0.0 and 0.25) may, in fact, bracket periastron. This could be consistent with the double passage of the neutron star through the circumstellar disk. In which case the third peak, currently labeled as falling at phase 0.55, moves closer to apastron. X-ray, but not optical activity, at this phase has previously been reported in other high mass X-ray binary systems - such as GX301-2 (Laycock et al., 2003) which has similar pulse and orbital periodicities to SXP327.

\begin{figure}
\begin{center}
\includegraphics[width=70mm,angle=-90]{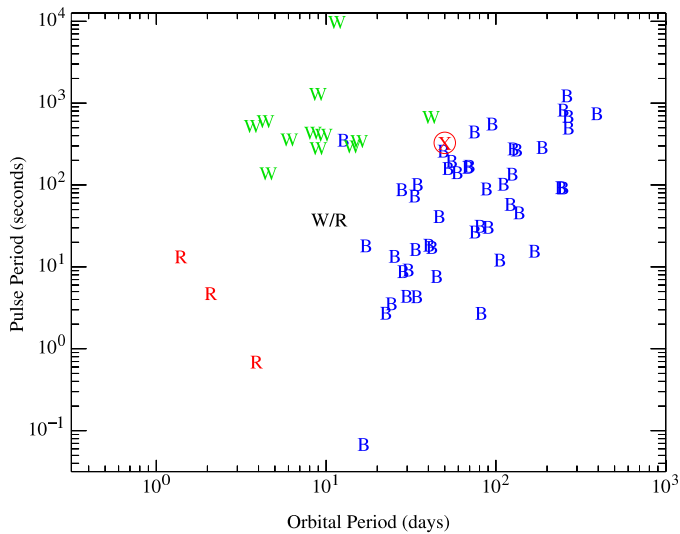}
\caption{Location of SXP327 (marked with a cross) on the Corbet diagram (modified from Corbet et al. 1999).The B symbol indicates
a Be star system, the W indicates a wind-fed system and the R symbol
indicates the Roche-lobe overflow systems.}
\label{fig:corbet}
\end{center}
\end{figure}

The location of SXP327 on the Corbet diagram is interesting - see Figure~\ref{fig:corbet}. The source falls on the outer fringes of the distribution of Be/X-ray binaries, quite close to the wind-fed hypergiant system, GX301-2 (Kaper et al., 1995) - the same source that also shows strong X-ray evidence for apastron emission. There is only one known supergiant system in the SMC, SMC X-1, so it would be of great interest to find another one. Though the OGLE III and MACHO data do not permit us to determine the true colours of this object, the average OGLE III I-band magnitude of 16.7 provides some useful information. Assuming a distance modulus to the SMC of 18.9, then a B1V star would have $m_{I}$ = 16.3 and a B2V star would have $m_{I}$ = 16.75. Whereas a supergiant B2I star would have $m_{I}$ = 12.7. Therefore there is an excellent support for the suggestion that this is another Be star with a spectral class that matches with the most common class of such systems in the SMC (McBride et al., 2008).

Finally, it is also noteworthy that there is an IR source, Sirius J00525223-7217151 (Kato et al., 2007), coincident with the optical counterpart with J=16.55$\pm$0.02, H=16.27$\pm$0.02 \& K=16.14$\pm$0.04. The observation was done on 3 July 2004. Hence, at that time,  J-K=0.41$\pm$0.04, which is similar to that seen from many of the previously identified SMC Be/X-ray binaries (Coe et al., 2005) and strongly indicative of emission from the circumstellar disk around a Be star.

\section{Conclusions}

Strong optical modulation is reported in the optical counterpart to a newly-discovered X-ray pulsar in the SMC - denoted here as SXP327. Though many systems reveal some degree of optical modulation at the binary period, the consistent degree of flaring seen here is exceptional amongst the group of $\ge$50 systems in the SMC. The optical period of 46d (presumably the binary period), and the X-ray pulse period of 327s, places SXP327 on the edge of the distribution of such objects on the Corbet diagram, but not far enough away to suggest it could be a rare supergiant system. In fact the optical magnitude supports the identification of this system as a new Be/X-ray binary in the SMC.

\section{Acknowledgements}

We are grateful to Richard Lamb for helpful comments.

This paper utilizes public domain data originally obtained by the MACHO Project, whose work was performed under the joint auspices of the U.S. Department of Energy, National Nuclear Security Administration by the University of California, Lawrence Livermore National Laboratory under contract No. W-7405-Eng-48, the National Science Foundation through the Center for Particle Astrophysics of the University of California under cooperative agreement AST-8809616, and the Mount Stromlo and Siding Spring Observatory, part of the Australian National University.

\bsp

\label{lastpage}

\end{document}